# VOCs monitoring using microwave capacitive resonator and conductive polymer – MWCNTs nanocomposites for environmental applications

P. Bahoumina[1*§], H. Hallil[1°§], J.L. Lachaud[1], A. Abdelghani[2], K. Frigui[2], S. Bila[2], D. Baillargeat[2], Q. Zhang[3,4], P. Coquet[3,4], C. Paragua[4], E. Pichonat[4], H. Happy[4], D. Rebière[1], C. Dejous[1§]

[1]Univ. Bordeaux, Bordeaux INP, CNRS, IMS, UMR 5218, F-33405 Talence, France
[2]Univ. Limoges/CNRS, XLIM, UMR 7252, F-87060 Limoges, France
[3]CINTRA, CNRS/NTU/THALES, UMI 3288, Singapore 637553, Singapore
[4]Univ. Lille, CNRS, IEMN, UMR 8520, F-59652 Villeneuve d'Ascq, France
(*) prince.bahoumina@ims-bordeaux.fr; (°) hamida.hallil@ims-bordeaux.fr, (§) *Member, IEEE*

*Abstract*— This paper presents a chemical microwave flexible sensor based on a resonant electromagnetic transducer in micro-strip technology with poly (3,4-ethylenedioxythiophene) polystyrene sulfonate – multi wall carbon nanotubes (PEDOT:PSS-MWCNTs) as sensitive material for Volatile Organic Compounds (VOCs) detection. The results show a high sensitivity to ethanol and toluene vapors on the S parameters of a passive resonator over a large frequency range. It is equal to -1.59 kHz/ppm and -1.45 kHz/ppm for ethanol and toluene vapors respectively. This kind of sensor can be integrated into real-time multi-sensing platform adaptable for the Internet of Things (IoT).

## I. Introduction

Atmospheric pollution involves several particles and mixture into complex gases, such as VOCs, which are detectable in the range of parts per million (ppm) for health environmental applications. Most of the commercially available sensors are based on conductivity transduction using metal oxides as a sensitive layer. However, such sensors often operate at high temperatures. On the contrary, electromagnetic transducers can operate at room temperature [1], which is just one of their advantages. To increase the sensitivity and to emphasize the selectivity, the active part is sometimes covered by a sensitive layer. In this study the sensitive material is nanocomposites of poly(3,4 ethylenedioxythiophene):poly(styrene sulfonate) ( PEDOT:PSS) and multi-wall carbon nanotubes (MWCNTs). The response of the sensor in a frequency range up to 6 GHz is presented, as well as the results of detection of ethanol and toluene vapors.

## II. Theoretical study

### A. Design

The device consists of two bandpass resonators in order to achieve differential detection. A resonator without any sensitive layer is considered as a reference channel, while the other resonator is the sensitive channel functionalized with a sensitive material to target the gas. Each channel consists of two parallel combs of 50 electrodes each. The geometry and configuration of the prototype is shown in figure 1, with a gap ($L_2$) between two successive electrodes of 300 μm.

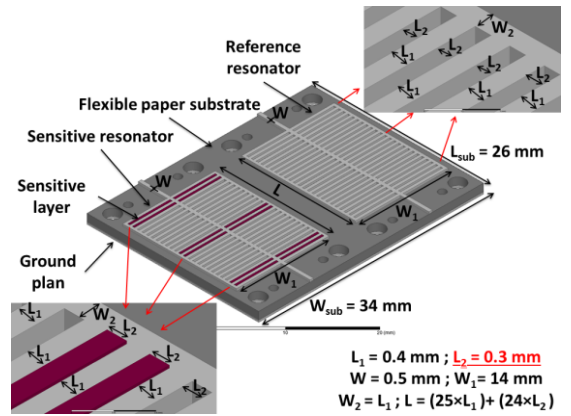

Figure 1. Pattern of the sensing dual device

### B. Simulation results

The operating principle of the sensor was achieved previously [1], [2]. In this study, we used a paper substrate with a permittivity of 3.05 and a dielectric loss of 0.089 at 4.3 GHz. Commercial inks based on silver nanoparticles and a conductive ink of PEDOT: PSS - MWCNTs were used as electrodes and as sensitive film, respectively. The material parameters were previously presented [2], [3]. The reference resonator can be used to compensate the variations not related to the sensitive material. Figure 2 shows the S parameters resulting from simulation of each channel. The first and second resonant mode of the reference and sensitive resonators are recorded around 2.5 and 4.85 GHz. The simulation of the presence of 5 layers of sensitive material induces a frequency shift close to -100 MHz for each mode. The distribution of the electric field E at both resonant modes of the reference channel is shown in figure 3. It puts to evidence areas with more intense level of E field, which have been selected for 12 bands of 5 sensitive layers placed in the 300 μm gaps of the measurement channel, as represented in

*Research supported by the French National Research Agency under project ANR-13-BS03-0010.

figure 1, in order to emphasize the interactions with the E field and so, a capacitive disturbance effect.

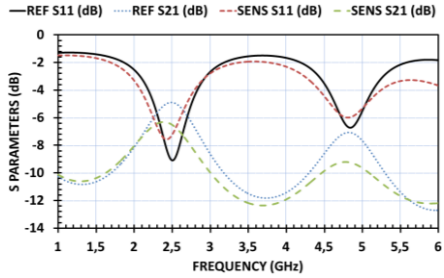

Figure 2. Simulated S parameters

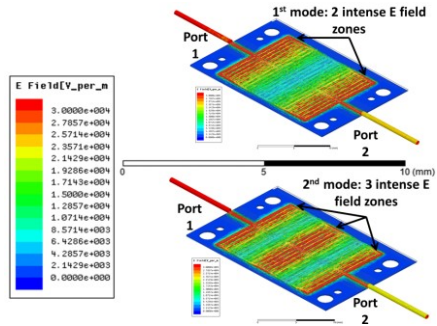

Figure 3. Distribution of the electric field E for the two resonant modes

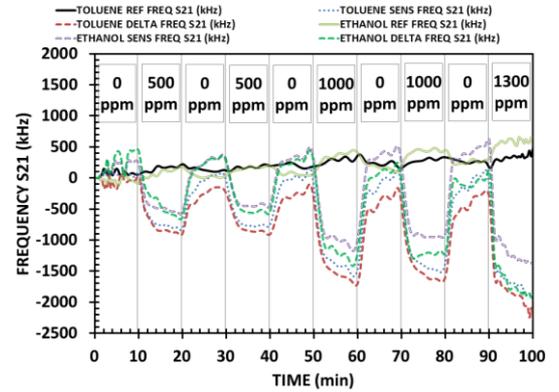

Figure 4. Real-time response to ethanol and toluene vapors

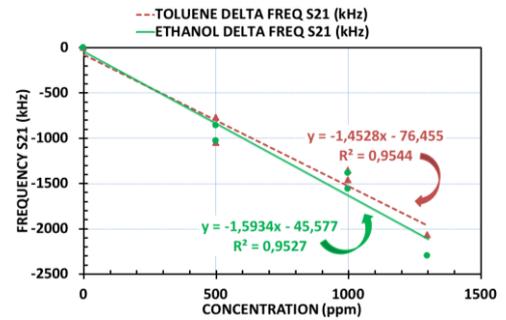

Figure 5. Estimation of the experimental sensitivity

III. EXPERIMENTAL STUDY AND DISCUSSION

The device was manufactured using the Dimatix inkjet printer (2800 series), the JSB 25 HV silver nanoparticles ink and the Poly-Ink HC ink (PEDOT: PSS-MWCNTs) as sensitive layer. The substrate was Epson photo paper with a thickness of 260 μm. The electrical characterizations were carried out with a vector network analyzer MS2026B, calibrated with 4001 points, over the frequency range up to 6 GHz for the general behavior of the device and in the frequency range from 2 to 4 GHz for vapors detection, focusing on the first resonant mode. The experimental detection configuration is mainly based on a vapors generator. The target vapors are transported by nitrogen as a carrier gas [2], [3]. Ethanol and toluene were used as target vapors, with concentrations up to 1300 ppm. The device was exposed to the vapors for 10 minutes each concentration step, carried out at room temperature. The 0 ppm concentration corresponds to nitrogen exposure only.

Figure 4 illustrates the effect of the ethanol and toluene vapors on the frequency of the resonators $S_{21}$ transmission parameters. The resonant frequency of the reference resonator increases slightly under exposure to the target vapors, while the frequency of the sensitive resonator decreases significantly. This is in accordance with an increase of the permittivity of the sensitive material. For a clearer understanding, the results are represented in differential detection mode, with $\Delta F\_S21$ obtained by subtracting the response of the reference from that of the sensitive channel. Thus we obtain the responses reported in figure 5, which show the evolution of $\Delta F\_S21$ for the device after 10 minutes of each concentration step. It is decreasing for both targeted vapors, making it possible to verify a coherent behavior, associated with an expected increase of the permittivity of the sensitive material. From these curves, it can be estimated that the sensitivity is equal to -1.59 kHz/ppm and -1.45 kHz/ppm for ethanol and toluene vapors, respectively.

IV. CONCLUSION

The feasibility of the capacitive microwave VOCs sensor has been demonstrated. We have shown that the ethanol and toluene vapors concentrations can be measured in real time detection. We can improve the sensitivity by increasing the sensitive surface, the thickness of the sensitive material, or by further functionalizing this layer, which could also address selectivity issue. A more thorough analytical study will improve the understanding of the physical and chemical phenomena that are involved.

ACKNOWLEDGMENT

The authors are grateful to the French National Research Agency for support under project ANR-13-BS03-0010 and the French RENATECH network (French National Nanofabrication Platform), as well as the French Embassy of Singapore (Merlion project).